# Single-molecule studies of the dynamics and interactions of bacterial OXPHOS complexes

Tchern Lenn[a] and Mark C. Leake[b,*]

[a] School of Biological and Chemical Sciences, Queen Mary University of London, Mile End Road, London, E1 4NS, UK.
[b] Biological Physics Sciences Institute (BPSI), Departments of Physics and Biology, University of York, York, YO10 5DD, UK

[*] Corresponding author. Tel: +44 1904 322697
Email address: mark.leake@york.ac.uk (M. Leake)

**Highlights**
- Bacterial OXPHOS can be explored at the single-molecule level using advanced fluorescence microscopy of fluorescent protein labelled cell strains
- Bacterial OXPHOS components are dynamic in the cell membrane
- OXPHOS supercomplexes, as evidenced in mitochondria, are not universally found in all bacteria

**Abstract**

Although significant insight has been gained into biochemical, genetic and structural features of oxidative phosphorylation (OXPHOS) at the single-enzyme level, relatively little was known of how the component complexes function together in time and space until recently. Several pioneering single-molecule studies have emerged over the last decade in particular, which have illuminated our knowledge of OXPHOS, most especially on model bacterial systems. Here, we discuss these recent findings of bacterial OXPHOS, many of which generate time-resolved information of the OXPHOS machinery with the native physiological context intact. These new investigations are transforming our knowledge of not only the molecular arrangement of OXPHOS components in live bacteria, but also of the way components dynamically interact with each other in a functional state. These new discoveries have important implications towards putative supercomplex formation in bacterial OXPHOS in particular.

## 1. Introduction

The enzymes and substrates used by bacteria for chemiosmotic ATP synthesis are as varied as the ecological niches they occupy. As autotrophs and/or heterotrophs living phototrophically, chemolithotrophically, and/or organotrophically, bacteria can use light, minerals and organic substrates as sources of reducing electrons for electron transport chains that establish a proton motive force (pmf) by redox reactions coupled to proton translocation, which can be vectoral (where protons are literally pumped through channels in the enzyme) or scalar (where protons are chemically consumed on one side of the membrane and liberated on the other). The pmf in turn is used to power the rotary mechanism of the $F_1F_o$-ATP synthase which produces ATP [1]. While there is great diversity in the enzymes and electron transport chains that allow bacteria to 'eat' everything from photons and water, to elemental sulphur and sugars, the basic mechanism of chemiosmotic ATP synthesis [2] is same and requires a pmf across a membrane, established by correctly oriented electron transport chain enzymes which effectively move protons across that membrane in the opposite direction to that which protons must flow through the ATP synthase for ATP synthesis.

Historically, the chemiosmotic ATP synthesis systems of the plant chloroplast and mammalian mitochondrion are the bases of understanding the two classical modes of chemiosmosis: photo-phosphorylation in the former and oxidative phosphorylation (OXPHOS) in the latter [3]. Figure 1 schematises chemiosmotic ATP synthesis by oxidative phosphorylation (OXPHOS) in the model heterotrophic bacterium *Escherichia coli*, illustrating the metabolic flexibility of this organism. (For more detail on *E. coli* OXPHOS genes, we refer the reader to the review by Magalon and Alberge, BBA 2015, in this special issue and figure 1 of that review in particular).

This central metabolic pathway is governed by a multi-enzyme system that is localized to bioenergetic membranes, but the organization and dynamics of bioenergetic complexes in two dimensions in the plane of the membrane is not well understood and has implications for the operation of the system as a whole. Two extreme models of organization and dynamics can be envisaged: solid state or random diffusion [4]. A solid state model implies that protein complexes are locked together and substrates are channelled from one to the other, such that the efficiency of the system would be limited by the turnover of the enzymes, whereas a random diffusion model allows for the possibility that the concentration of components in the membrane limits flux.

The presence of oxidative OXPHOS supercomplexes in mitochondria and bacteria [5-10] and role of supercomplexing for channeling electrons [11] support of a solid state model in mitochondria, but the relevance of supercomplexing to the catalytic kinetics and efficiency of the system is disputed by alternative interpretations [12] of the data of Lapuente-Braun et al [11]. Blaza et al [12] in fact suggest that while supercomplexes exist in mitochondria, perhaps they have no physiological function other than to allow optimal enzyme packing and thus improve overall efficiency of OXPHOS at the level of the whole mitochondrion.

What is the situation in bacteria? The operation of bacterial electron transport chains should be of great importance for those interested in killing or manipulating bacteria for disease management, bioproduction and bioremediation. For instance, altering levels of mobile electron carriers might have large effects for bacteria which operate a random collision system, but might have little effect on bacteria that operate at solid state system.

Fluorescence microscopy studies have allowed researchers to address the questions of the organization and dynamics of bacterial OXPHOS components in live cells. Traditional fluorescence imaging techniques involve exciting and collecting the emitted fluorescence from all the fluorophores in the focal plane simultaneously. Images are generated directly by photons landing on a 2-dimensional detector, which translates the signal into a photon intensity map – or takes a photograph (confocal images can be considered to be a photo-collage). Such images are diffraction-limited in terms of spatial resolution to the typical optical resolution limit of ca. 200-300 nm which is determined from the Abbe theory of optical diffraction to be roughly half the wavelength of emitted light. These diffraction-limited fluorescence imaging techniques suggested a heterogeneous distribution of OXPHOS complexes in the plasma membranes of *Bacillus subtilis* [13] and *Escherichia coli* [14] and that OXPHOS complexes are mobile in the bacterial membrane.

Single-molecule imaging approaches aim to build up a picture of the cell by observing many molecules but individually, one at a time. Such approaches not only reveal the overall trend for a population of molecules but also the population structure, showing up heterogeneities that may be averaged out in an ensemble measurement. They also have the advantage of being able to determine the location of molecules 10-50-fold more accurately than in diffraction-limited imaging [15]. This is because the detected emission from a point source of light on a two-dimensional detector is manifest as a point spread function image of 200-300 nm width which can be fitted by an analytical function (typically approximated by a 2D Gaussian profile) to determine the location of the intensity centroid as the best estimate for the location of that source [16]. Such studies on OXPHOS in bacteria have painted a detailed picture of bacterial OXPHOS systems in live cells, which is suggestive of how the many enzymes might work together to achieve ATP synthesis.

As far as we are aware, single-molecule fluorescence studies on bacterial OXPHOS have only been carried out on *E. coli* [17-20]. These studies have characterized the patches of OXPHOS complexes that have been observed by ensemble average imaging to unprecedented levels of detail. They have also catalogued the mobility of complexes in the membrane, taking advantage of the improved spatial resolution of single molecule approaches, revealing that the movement of complexes is not uniform. They have thus begun to address the question of how OXPHOS complexes relate to each other spatially.

**2. Heterogeneous Patching**

To date, the quinone reducing enzymes: type 1 NADH dehydrogenase (NDH-1) [21, 22] and succinate dehydrogenase (SDH) [23], quinol oxidising enzymes: the Cytochrome bo [24] and Cytochrome bd-1 complexes [25-27], and the $F_1F_o$-ATP synthase [28, 29] have been functionally fluorescent-protein labelled and expressed from native loci in *E. coli* cells (Table 1). By ensemble average imaging, all of these complexes were observed to be heterogeneously distributed in the *E. coli* membrane and apparently localized in mobile clusters [17, 30]. A more precise single-molecule approach was taken to study these apparent clusters in order to tease out details such at the variation in the number of complexes in each cluster, the distribution of physical sizes of the clusters and the diffusional behaviour of individual clusters or complexes [17, 19].

Single-molecule counting methods (see Box 1), developed originally from stoichiometry studies of torque-generating components of the bacterial flageller motor [31], revealed that these clusters are heterogeneous in terms of the number of complexes located within them (Table 2). The clusters are also expected to be variable in size as estimated by the comparison of the apparent full width at half maximum (FWHM) of the individual clusters to that of a single fluorophore in the same microscope [17].

PALM (**P**hoto-**A**ctivation **L**ocalisation **M**icroscopy) imaging of Cytochrome bd-1 and NDH-1 in fixed cells gives a more detailed picture of the arrangement of these complexes in the membrane [19], see Box 2. Consistent with the broad distributions that were suggested by diffraction-limited imaging, the clusters of complexes were inconsistent in shape and population and appear randomly located in the plane of the membrane. Single complexes were interspersed between clusters confirming that the apparent clusters of proteins observed in diffraction-limited images are in fact rather loose associations of proteins rather than rigid structures. In this case of fixed cells, proteins were therefore assumed to be immobile, with complexes functionally tagged with the photo-switchable fluorescent protein mMaple [32]. Localizations of mMaple were calculated according to Lee et al. (2012) [33] and the images were rendered similarly to Betzig et al. (2006) [34] and interpreted qualitatively.

**3. Two modes of Diffusion**

Additional evidence for the patchwork organisation of OXPHOS complexes comes from observing the diffusional behavior of single OXPHOS complexes in the *E. coli* membrane. The mean squared displacement (MSD) *vs* time interval relation of a proportion of tracked fluorescent spots in [17] and [19] plateau, indicating confined diffusion of molecules within a membrane domain of approximately 100 nm in diameter. Renz et al. (2012) [18] suggest that this is an artefact of a small imaging window due to the curvature of the cell, however, Renz et al. (2012) [18] inconsistently report the width of the short axis of an *E. coli* cell, reporting 500 nm in the discussion, while the total internal reflection fluorescence (TIRF) micrograph shown with 1 μm scale bar seems to indicate a width closer to 1 μm. Also, the diffusion coefficient D used in the simulation data that demonstrates that diffusion perpendicular to the long axis of the cell is systematically underestimated, is approximately twice that calculated for the real data. The conclusions of the paper would be more convincing if the authors had shown window size dependency of calculated D over a range of diffusion coefficients as the authors themselves state that the 'size limit should increase with faster diffusion'. Conversely the size limit should decrease with slower diffusion - i.e. the diffusion coefficient of proteins in an *E. coli* cell might not be underestimated estimation for slow moving objects (D < 0.18 $\mu m^2$/s). In fact, for Llorente-Garcia et al. [19], the complexes observed had a mean D of 0.007 $\mu m^2$/s and the authors report similar dimensional diffusion coefficient perpendicular or parallel to the long axis of the cells. The Bayesian ranking of diffusion (BARD) analysis [35] (see Box 3) used to classify tracks in Llorente-Garcia et al., [19] is a proposed solution to the problem of short tracks mentioned in Renz et al., 2012 [18]. Llorente-Garcia et al., [19] detected both classes of tracks with MSD *vs* time plots that did plateau (i.e. putative confined diffusion) and tracks that did not (i.e. putative free Brownian diffusion). If the confined diffusion was an artifact, surely all tracks would appear confined. Finally, it is worth noting that Llorente-Garcia et al. [19] only used tracks that contained 5 consecutive data points (i.e. could be tracked for at least 5 frames) whereas Renz et al. 2012 [18] seems to have included shorter tracks. This may be the reason for the discrepancy in D as selecting only longer tracks may have biased sampling to slow moving complexes as fast moving complexes may have moved out of the field of view within 4 frames. Nonetheless free and confined diffusion of OXPHOS complexes have been observed in *E. coli*.

Renz et al 2015 [20] report PALM and single particle tracking (SPT) PALM imaging of the ATP synthase in live *E. coli*. This is an insightful study, but unfortunately the authors do not make clear how the PALM images shown are rendered and therefore it is unclear as to whether the images shown take into account multiple localizations of the same molecule that has moved between image frames. Nonetheless, the presence of label-dense regions at a sample temperature of 37 °C is suggestive of the clusters described in Erhardt et al. [30]. The authors also report valuable work into the counting of complexes, but unfortunately the authors do not make clear how they solved the problem of over-counting [33, 36, 37] due to blinking, reactivation, long-lived fluorophores and movement (which is particularly to be expected in this case). The fluorophore used in this study, mEos3.2, is known to blink [38] so this is a caveat to be considered in interpreting the localization data. Their report of the diffusion coefficient of the ATP synthase is an order of magnitude above that reported by Llorente-Garcia et al. [19]. The discrepancy may be due to temperature differences, or differences in the genetic background of the strains. It is also notable that cells were grown at different temperatures in the two studies (30 °C for Llorente-Garcia et al [19] and 37 °C for Renz et al. 2015 [20]) which may affect membrane fluidity and phase transition temperature [39]. There is no presentation of the distribution of calculated diffusion coefficients or consideration of the possibility of confinement, which in effect reduces the diffusion analysis to a useful ensemble average study, rather than exploiting the possibility of revealing interesting heterogeneities in the population of tracks.

**4. Co-localization of OXPHOS components, or not?**

The prevalence of clustering of the labeled OXPHOS complexes lead to the hypothesis that even though supercomplexes of these enzymes had not been found, they may be corralled into patches in the *E. coli* membrane, dedicated to OXPHOS, which effectively improves the efficiency of electron transfer [40].

This hypothesis was tested by multi-color imaging of strains where one complex was labelled with a protein with red fluorescence and another complex was labelled with green fluorescent protein. Multi-colour imaging showed that it was unlikely that unlike complexes were corralled together in patches of membrane dedicated to performing OXPHOS [19].

Drawing this conclusion was not trivial from the available data because consideration of spatial and temporal scale is important when defining co-localization. It would be true to say that all OXPHOS complexes are co-localized within the bacterial cell, but co-localization is irrelevant on the micrometer scale of a whole bacterial cell, even though co-localization on a similar spatial scale could be relevant for if the question at hand was whether or not two proteins are located within the same organelle in a eukaryotic cell.

In the case of *E. coli* OXPHOS, the respiratory islands were estimated to be roughly 100 nm in diameter based on the FHWM of observed clusters [17]; the authors were therefore interested in co-localization on this scale, which is below the resolution limit for optical microscopes. The minimum proximity with which 2 objects can be confidently co-localized by direct observation in optical microscopy is given by the point spread function width of the microscope, or typically 200-300 nm - i.e. two subunits unit of a small protein complex that are in contact with each other can appear just as 'co-localized' as two proteins that are independently floating within a round corral that is 300 nm in diameter. For mobile objects, timescale is also an important consideration as one would want to distinguish between objects that have independently explored the same area in the field of view within the temporal observation window, and objects that are stably associated over time.

In Llorente-Garcia et al. [19], images were recorded with a 40 ms camera exposure time (this being the temporal resolution in this study) and co-localization of OXPHOS complexes was considered on various time scales.

On the time scale of seconds, co-localization was measured by analyzing frame averaged images of 25 consecutive image frames. These images revealed the aforementioned immobile/slow-moving patches as bright regions in the field of view. Qualitative analysis of the images clearly showed that the immobile/slow-moving patches of the two complexes observed in each cell strain tended to be located in different parts of the cell. A more quantitative analysis measured co-localization in approximately 20% of the pixels and a similar result was obtained on a 40 ms (i.e. single image frame) time scale (see Box 4). However these metrics were obtained in an ensemble imaging approach - limited by the optical resolution of the microscope.

A single-molecule approach was also taken, by allowing the sample to photobleach until single diffusing spots in each channel were seen. In sections of the video where single spots could be simultaneously tracked in both channels for at least 5 image frames, the overlap integral of fluorescent spots (approximated as the overlap of two 2D Gaussian intensity profiles) was calculated at each time point. Co-localization was defined as an overlap score of at least 0.2, according to the Raleigh resolution criterion. The percentage of time points where spots were co-localized was less than but close to the frequency predicted by chance proximity due to random walking. Co-localization events were therefore not frequent enough to be reasonably due to anything more than 'chance meetings' of randomly diffusing complexes in the cell membrane.

The authors conclude that NDH-1 is not co-localized with Cytochrome bd-1 nor SDH and Cytochrome bd-1 does not co-localize with ATP synthase, except when, apparently by chance, two complexes, or two patches of complexes, happen to drift close to each other. There are no data for combinations of NDH-1 and Cytochrome bo, nor ATP synthase with any other complexes. Unfortunately the double-labelled cell mutants tested did not include combinations of OXPHOS supercomplexes which have been suggested in *E. coli* [10, 41, 42].

## 5. Summary and Further questions

We observe: 1) the existence of patches of complexes; 2) a lack of co-localization of the pairs of complexes observed; 3) a fluorescent ubiquinone analog, NBDHA-Q, is not patchy in the membrane and diffuses much faster than OXPHOS complexes; 4) the oxygen consumption rate scaled in proportion to the diffusion coefficient of NBDHA-Q [19]. These observations suggest that the rate of electron transport through OXPHOS is limited not by the slow mobility of the membrane patches in *E.*

*coli*, but rather by the mobility of the delocalized NBDHA-Q. In other words, that the entire cell membrane should be considered to be a single electron-transport compartment, rather than a patchwork of closed electron transport circuits. While OXPHOS supercomplexes have been suggested from biochemical data in *E. coli* [10, 41, 42], the physiological function of these supercomplexes and the metabolic flux through them *in vivo* is as yet unclear. Currently, random collision accounts better for the mechanism of electron transport in *E. coli* than a solid state model, see Fig. 2.

What about protons? *In vivo* 2-color imaging of the ATP synthase and SDH in *B. subtilis* [13] did not suggest co-localization and neither was Cytochrome *bd-1* co-localized with ATP synthase in *E. coli* [19]. There is no *in vivo* evidence for localized proton circuits. However, the ATP synthase activity, along with NADH oxidation activity, was found to be higher in membrane vesicles containing flagella, than unflagellated vesicles - there was apparent Förster Resonance Energy Transfer between fluorescently labelled flagellar motor components and subunits of these OXPHOS complexes, however the authors do not demonstrate that the fluorescently labelled peptides are incorporated into active complexes [43]. Nonetheless, the immobile/slow moving spots of OXPHOS complexes observed by Llorente-Garcia et al. [19] may be complexes associated with the flagellar motor, which is also immobile in the membrane - providing a local 'power pack'. (But note, a dual-fluorophore labelled NDH-1-ATP synthase mutant has not yet been reported). Local pmf-power plants, close to proton-consuming processes is an attractive hypothesis but is yet to be proven.

Single-molecule fluorescence studies of OXPHOS in bacteria have concentrated on the genetically tractable model *E. coli*. They show that in this bacterium, OXPHOS complexes tend to patch in the membrane but only with like-complexes. Given that ubiquinol appears to be uniformly distributed and is fast diffusing, electron transport in *E. coli* is likely occur by random collisions of freely diffusing quinone/quinol molecules with relatively slow moving OXPHOS enzymes that do not tend to mingle.

These single-molecule studies have also given us a unique view of the *E. coli* membrane, which appears to be a patchwork rather than a homogenous mixture. How this patchwork arises is still a mystery, as is its physiological significance, if any. Is the explanation of Blaza et al. [12] for supercomplexing in mitochondria also applicable to the *E. coli* membrane? Is patching the evolutionary solution to optimally pack the membrane with apparatus for energy production and other metabolic processes, transport, secretion, motility, cell division and sensing? Is this patchiness of proteins truly a generic feature of bacterial membrane architecture and what are the implications for membrane function?

From an ecological/evolutionary perspective, it is also interesting to ask if there is a fitness advantage for *E. coli* to have OXPHOS complexes in patches rather than dispersed and while disruption in supercomplex formation is seen in mitochondrial dysfunction [44], how important are supercomplexes for bacterial OXPHOS?

Rich veins of research have been opened up by single-molecule imaging of bacterial OXPHOS. Quantitatively considering the organisation and dynamics of OXPHOS components in space and time provide excellent examples of experimental single-molecule cellular biophysics and single-molecule cell biology in practice. These experiments are a good demonstration of just how far light microscopy has come since its inception over 300 years ago [15, 16, 45-67].

**Acknowledgements**

M.C.L was funded in part by the Biological Physical Sciences Institute (BPSI) at York University and from a Royal Society University Research Fellowship.

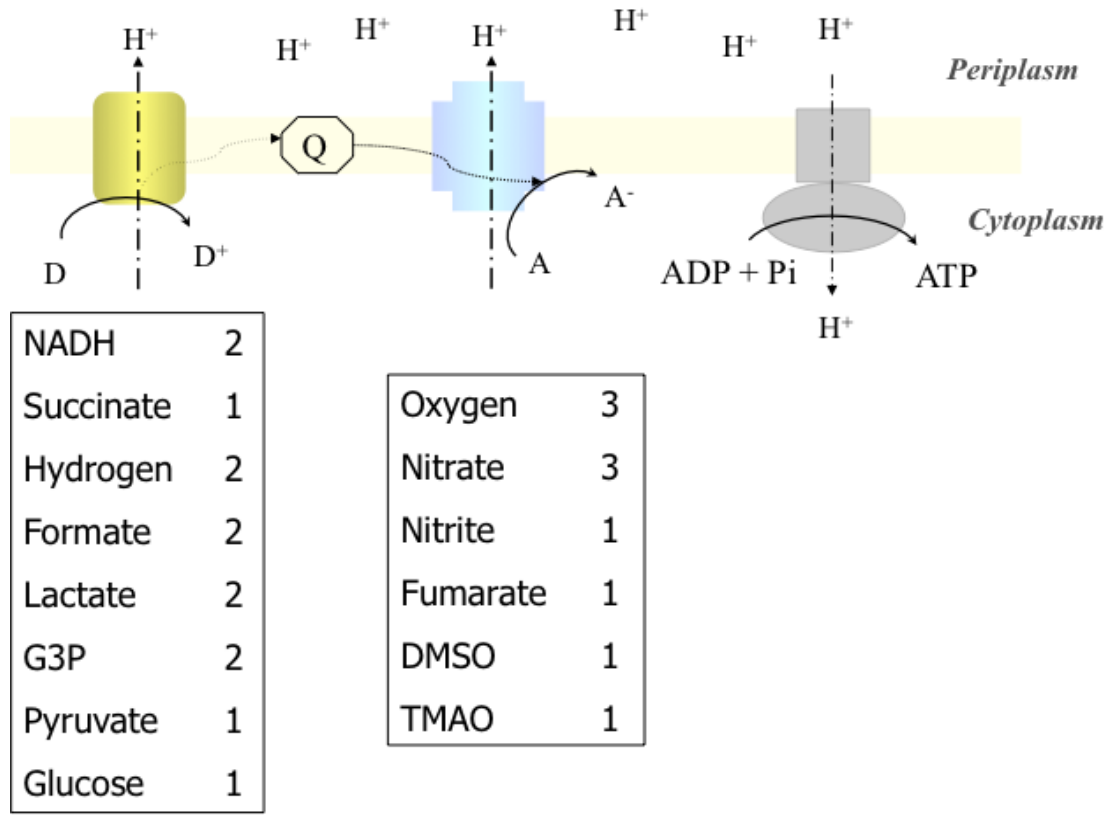

Figure 1: **Schematic representation of the *Escherichia coli* OXPHOS chain**

Electrons are transferred from donor substrates (D) to quinones (Q) by quinone reductases (in yellow) and then to acceptor substrates (A) by terminal oxidases (in blue). Electron transport is concomitant with proton translocation, resulting in a pmf. for chemiosmotic ATP synthesis by the $F_1F_O$ ATP synthase (in grey). Below each depicted enzyme is the range of electron donors and acceptors that can be used by the *E. coli* respiratory chain and the numbers indicate the number of known isozymes for each substrate (data adapted from [68]).

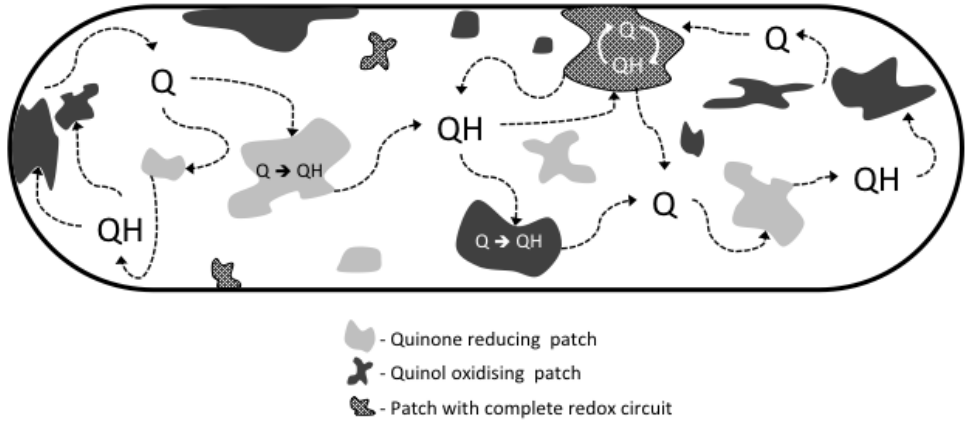

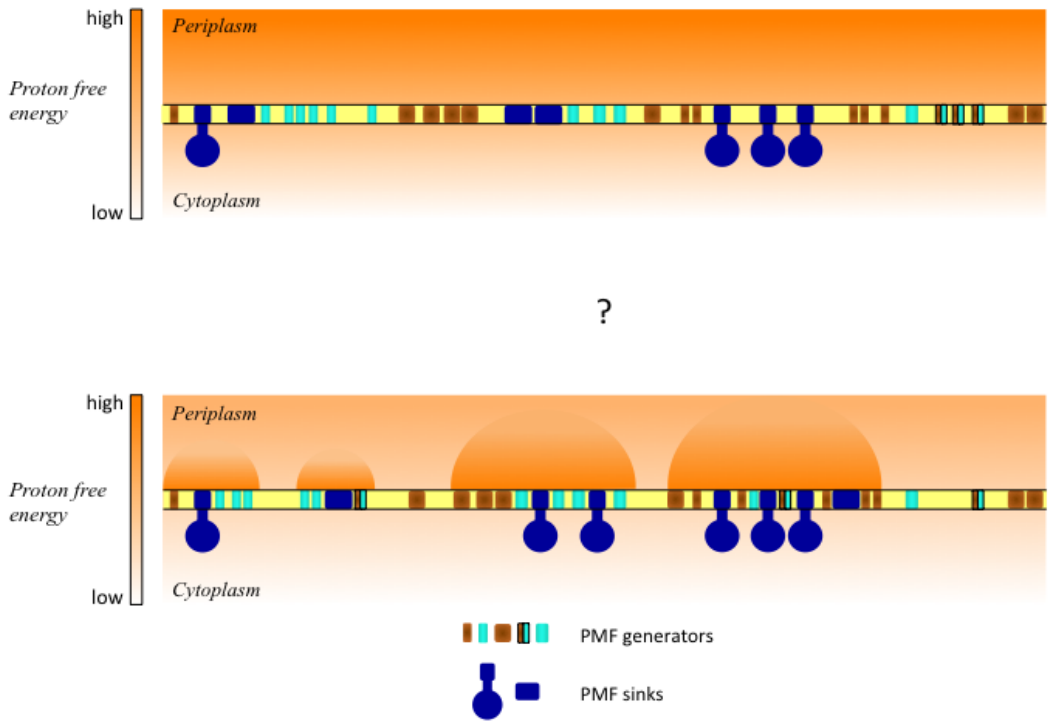

**Figure 2: Models of Electron Flux (A) and the PMF (B) in *E. coli*.**
- A) While patches of OXPHOS complexes exist alongside individual enzymes (not represented) it is likely that each patch contains only one type of OXPHOS enzyme, either reducing quinone (Q) to quinol (QH) or oxidizing QH to Q, which freely diffuse in the membrane plane. While supercomplexes of complete redox circuits have been detected *in vitro* [42], these have yet to be studied *in vivo*.
- B) It is unclear, as yet as to whether the 'proton pool' is effectively uniform as the 'quinone pool' appears to be in a *E. coli* (upper panel) or if 'proton sinks' have 'local power supplies' (lower panel).

**Box 1: Single-molecule counting methods (developed from [31]):**

- *Step-wise photobleaching:*
    o Dye molecules (e.g. used as reporter molecules, such as fluorescence proteins (FPs)) photobleach irreversibly
    o Size of the photobleach step is a 'molecular signature' equivalent to the brightness of a single dye molecule unique to the physical and chemical environment of that molecule
    o Steps, for example multiple steps from a molecular complex tagged with multiple dye molecules can be detected directly in the 'time domain', but this is error-prone since it is difficult to objectify the detection of 'real' steps over noise
    o A more robust approach involves Fourier spectral analysis, involving generating the pairwise difference distribution of fluorescent spot intensity values and calculating the power spectrum of this to estimate the fundamental peak which is the unitary brightness of one dye molecule, $I_{FP}$.
- *Calibration with single FPs in vitro:*
    o Single FP molecule can be immobilized to a coverslip using anti-FP antibodies for example.
    o The brightness of each molecule can be measured under the same imaging conditions but using a more controlled chemical environment.
    o The brightness of single FP molecule *in vivo* may be different from the in vitro value due to differences in local pH and halide ion concentration, in addition to some small local differences in laser excitation intensity.
    o Experiments on model systems suggest that under most circumstances the *in vivo* and *in vitro* brightness values agree to within ~10%.
    o This allows the *in vitro* value to be used as a guide on the sometimes noisy power spectrum from the *in vivo* Fourier special analysis to facilitate finding the correct peak corresponding to the single-molecule FP brightness.
- *Calculation of stoichiometry:*
    o The initial brightness $I_0$ of a fluorescently-labeled molecular complex can be estimated from an exponential fit to the spot intensity *vs* time data during photobleaching observed from a given tracked spot.
    o The stoichiometry for each spot can be estimated as $I_0/I_{FP}$.
    o The error on the stoichiometry estimate is sub single-molecule for typical values of stoichiometry of 10 molecules or less per spot.
    o For higher stoichiometry spots the error in the stoichiometry estimate is greater than a single molecule for a given single spot, however performing this estimate on several such spots allows estimates for mean values in a population of molecules, or even in sub-populations if identified as distinct peak in the stoichiometry probability distribution, to be made which have better than single-molecule precision.

**Box 2: PALM microscopy:** *E. coli* **cells expressing mMaple labeled Cytochrome bd-1 oxidase.**

These PALM images are rendered such that information about the intensity centroid estimate for each localized molecule and the degree of certainty about the location that centroid are taken into account. Each photon burst, which meets various selection criteria, such as brightness, duration and area, is considered to be a localization event of a single molecule and its centroid determined by the location of the peak of its point-spread function. Each localized molecule is rendered as a 2D Gaussian with normalised integrated volume, centered on the calculated intensity centroid with width determined by the uncertainty of centroid localization, which is dependent on the number of photons collected in that event [69, 70]. Variations in the uncertainty of centroid location can be due to:
- Brightness of individual events – the centroids of bright localizations can be more accurately located than dim ones [69, 70].
- Duration of localization event – more photons will be collected for long-lived molecule that "resists" photobleaching.
- Variation in camera noise and autofluorescent background [69].
- Movement of molecules [71] (even in fixed samples [72]).

In this rendering, the final image is the sum of rendered localizations and could therefore be interpreted as a probability map of molecule locations where the brightest areas correspond to the regions where the reader can be most confident of where molecules are located and the dimmest areas are highly unlikely to contain any labeled complexes – assuming that the labeled complexes are well sampled.

**Box 2 Figure:** PALM images of *E. coli* cells expressing mMaple tagged Cytochrome bd-1 [19].

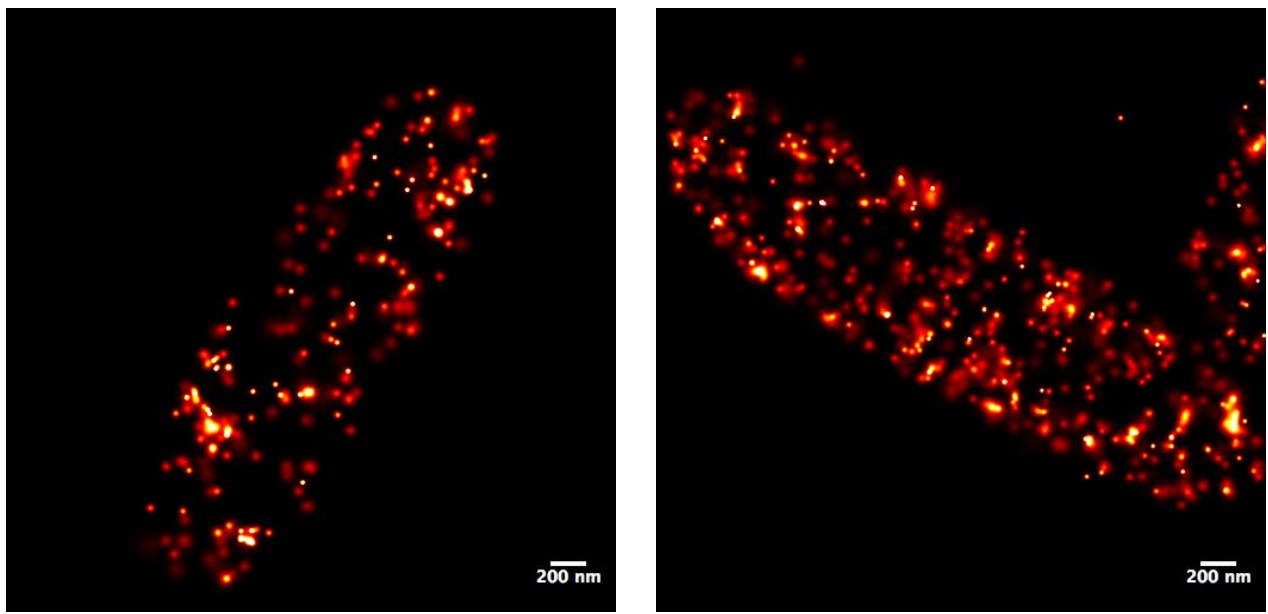

**Box 3: SPT and BARD analysis**

*Single Particle Tracking:*
- Assumes objects in sequential fields of view that have similar characteristics to each other (shape, intensity) and do not differ in position beyond a pre-selected distance are the same object which has moved in between frames.
- The path of the object takes in two dimensions can be reconstructed by plotting the position of the identified object over time.
- The object is thus 'tracked' and various parameters such as path length and mean-squared displacement can be calculated.
- When multiple objects are present in the same frame, the complexity of the tracking problem increases, especially if the objects are very similar.

*BARD analysis* [35]*:*
- Various modes of diffusion, for example:
    - Free (Brownian)
    - Confined
    - Anomalous
    - Directed
- Assignation of diffusion mode is problematic for short tracks typical of fluorescent protein labeled cell strain
- Objective Bayesian inference is used to predict which diffusion mode from a given list is the one which best accounts for the observed SPT data
- Algorithm can be expanded into far more complex heterogeneous diffusion modes e.g. binding/release events

**Box 4: Two co-localization metrics in Llorente-Garcia et al [19] :**

- Multiple length scale methods, can infer co-localization:
    - Cell-by-cell
    - Track by track
    - Pixel-by-pixel
- Multiple time scale methods, can infer co-localization on separate scales of:
    - 10s of seconds
    - Seconds
    - 10s of milliseconds
- Pixel-level statistics:
    - Normalize intensity values in each separate color channel min-max
    - Look at normalized ratio of intensity values of corresponding pixels in each color channel
    - Pixels where only 1 fluorophore is present will take on extreme values (1 or -1)
    - Pixels containing signals from both fluorophores (or none) will be close to 0
- Gaussian overlap method:
    - Models each fluorescent spot as a 2D Gaussian intensity profile
    - Tracks probed from each color channels which are coincident in time
    - Overlap integral calculated on the basis of integrated overlap between these 2D Gaussians as a metric for co-localization (or not)

**Table 1: Funtionally labelled *E. coli* OXPHOS complexes**

| OXPHOS Complex | Subunit | Terminus | Linker | Fluorophore | Reference source |
|---|---|---|---|---|---|
| NDH-1 | NuoF | N | Thr-Asp-Pro-Ala-Leu-Arg-Ala | GFP, mCherry, mMaple | [30], [17],[19] |
| NDH-1 | NuoF | C | Gly-Leu-Cys-Gly-Arg | cerulean | [30] |
| SDH | SdhC | N | Thr-Asp-Pro-Ala-Leu-Arg-Ser[*] | mCherry | [30] |
| Cytochrome bd-1 | CydB | C | Gly-Leu-Cys-Gly-Arg | GFP, mCherry, mMaple | [17], [19] |
| Cytochrome bo | CyoA | C | No linker | mCherry | [30] |
| ATP-synthase | AtpB | C | Gly-Ser-Met-Val | GFP | [30] |
| ATP-synthase | AtpB | C | Gly-Ser | GFP, PAGFP, mEos3.2 | [20] |

[*]replaces Met-Ile-Arg-Asn

**Table 2**: Stoichiometry of OXPHOS enzyme clusters

| OXPHOS Complex | Range of number of complexes per cluster |
|---|---|
| NDH-1 | 10-20 |
| Cyt bo | 24-45 |
| Cyt bd-1 | 70-180 |
| SDH | 20-40 |
| ATPase | 40-60 |